\documentclass[10pt,twocolumn,letterpaper]{article}

\usepackage[pagenumbers]{cvpr} 

\usepackage[dvipsnames]{xcolor}

\definecolor{cvprblue}{rgb}{0.21,0.49,0.74}
\usepackage[pagebackref,breaklinks,colorlinks,citecolor=cvprblue]{hyperref}
\usepackage{multirow}
\usepackage{csquotes}
\newcommand{\ours}{{\sc GR-Gaussian}\xspace}

\usepackage{algorithm}
\usepackage{algorithmic}

\usepackage{color}
\definecolor{color3}{rgb}{0.95,0.95,0.95}
\definecolor{Red}{RGB}{192, 0, 0}
\definecolor{Blue}{RGB}{12, 114, 186}
\definecolor{Yellow}{RGB}{218, 169, 20}

\def\paperID{*****} 
\def\confName{CVPR}
\def\confYear{2024}

\title{\ours: Graph-Based Radiative Gaussian Splatting for Sparse-View CT Reconstruction}

\author{Yikuang Yuluo$^{1, *}$\qquad\quad Yue Ma$^{2, \dagger}$\qquad\quad Kuan Shen$^{1}$ \qquad\quad Tongtong Jin$^1$ \\ Yangpu Ma$^1$\qquad\quad Wang Liao$^1$ \qquad\quad   Fuquan Wang$^{1, \dagger}$\\
$^1$ Chongqing University \qquad $^2$ HKUST \\
\url{}
}

\begin{document}
\twocolumn[{
\maketitle

\begin{center}
\captionsetup{type=figure}
\includegraphics[width=\textwidth]{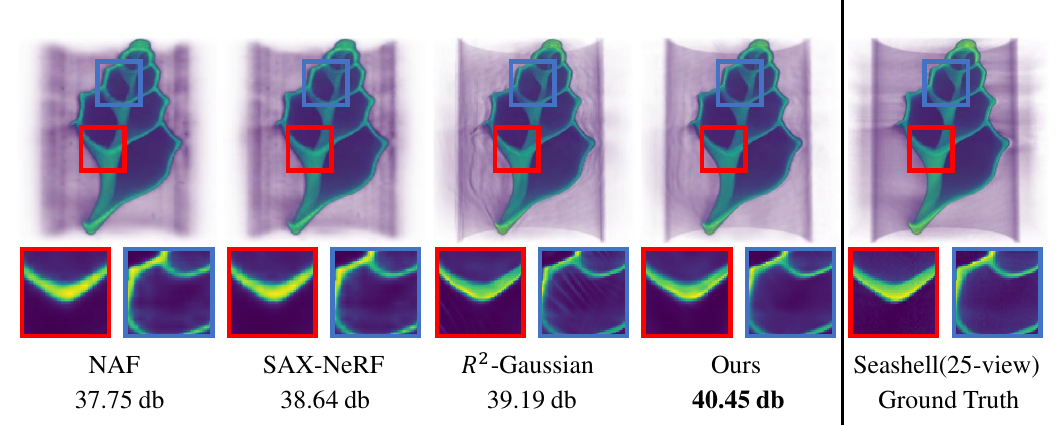}
\captionof{figure}{
Visual results of \ours. We compare \ours to two NeRF-based methods NAF\cite{zha2022naf}, SAX-NeRF\cite{cai2024structure}) and $R^2$-GS\cite{r2_gaussian} in terms of visual quality and PSNR (dB). Our method mitigates needle-like artifacts and achieves superior CT reconstruction quality under sparse-view conditions.
}
\label{fig:title case}
\end{center}
}]

\renewcommand{\thefootnote}{\fnsymbol{footnote}}
\footnotetext{$*$~Equal contribution.}
\footnotetext{$\dagger$~Corresponding Authors.}

\begin{abstract}
3D Gaussian Splatting (3DGS) has emerged as a promising approach for CT reconstruction. However, existing methods rely on the average gradient magnitude of points within the view, often leading to severe needle-like artifacts under sparse-view conditions. To address this challenge, we propose GR-Gaussian, a graph-based 3D Gaussian Splatting framework that suppresses needle-like artifacts and improves reconstruction accuracy under sparse-view conditions. Our framework introduces two key innovations: (1) a Denoised Point Cloud Initialization Strategy that reduces initialization errors and accelerates convergence; and (2) a Pixel-Graph-Aware Gradient Strategy that refines gradient computation using graph-based density differences, improving splitting accuracy and density representation. Experiments on X-3D and real-world datasets validate the effectiveness of GR-Gaussian, achieving PSNR improvements of \textbf{0.67 dB} and \textbf{0.92 dB}, and SSIM gains of \textbf{0.011} and \textbf{0.021}. These results highlight the applicability of GR-Gaussian for accurate CT reconstruction under challenging sparse-view conditions.
\end{abstract}    
\section{Introduction}
\label{sec:intro}
\begin{figure*}[tb]
  \centering
  \includegraphics[width=\linewidth]{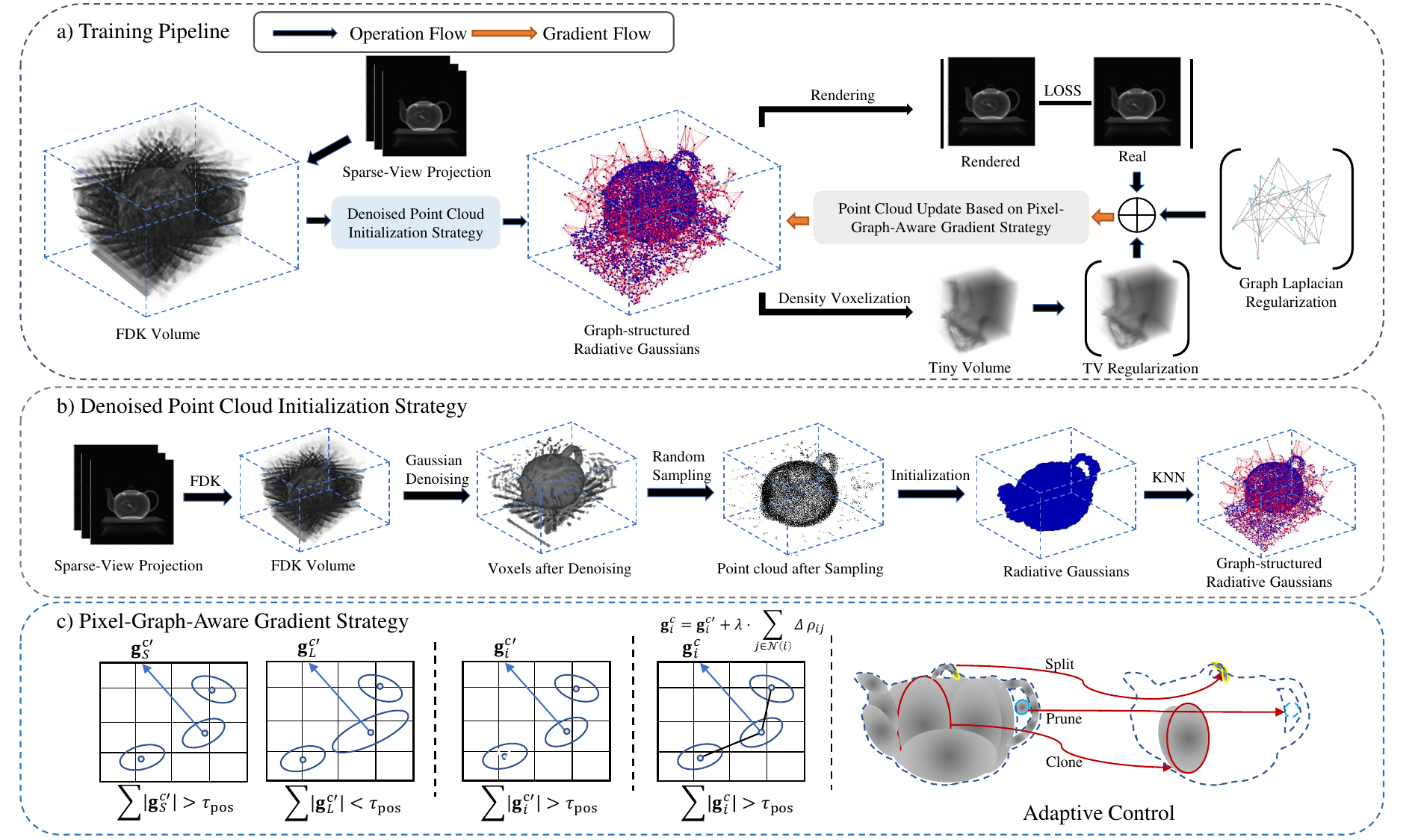}
  \caption{\textbf{Training pipeline of GR-Gaussian}. (a) Overall training pipeline. (b) Denoised Point Cloud Initialization Strategy. (c) Pixel-Graph-Aware Gradient Strategy.} 
  \label{pipeline}
\end{figure*}
ake the reconstruction problem highly ill-posed. Traditional analytical methods, such as FBP and FDK\cite{feldkamp1984practical}, are computationally efficient but often fail under these conditions, producing artifacts. Model-based optimization approaches\cite{andersen1984simultaneous,ma2024followyouremoji}, which use nonlinear regularizers and iterative solvers (e.g., ART), achieve better results but are computationally expensive.

Deep learning-based methods\cite{lin2023learning,chung2023solving,liu2023dolce} demonstrate strong performance but require large labeled datasets and long training times\cite{ma2025followcreation,ma2025followyourmotion}. They also struggle with out-of-distribution objects. NeRF-based\cite{mildenhall2021nerf,ma2025followyourclick,ma2024followpose} approaches show promise in per-case reconstruction but are extremely time-consuming due to extensive point sampling during volume rendering\cite{zha2022naf,cai2024structure,zang2021intratomo}. These methods highlight the trade-off between reconstruction quality and computational efficiency. 

Recently, 3D Gaussian Splatting (3DGS)\cite{kerbl20233d,yu2024mip,guedon2024sugar} has demonstrated superior quality and efficiency over NeRF in view synthesis tasks\cite{lu2024scaffold,liang2024gs}. However, when applied to sparse-view tomographic reconstruction, 3DGS-based methods exhibit characteristic needle-like artifacts. Through analysis, we identify that these artifacts arise from the retention of large Gaussian kernels with small gradients during the reconstruction process. This limitation stems from the lack of consideration for inter-point relationships, which results in biased density distributions and insufficient splitting of Gaussian kernels.

To address these issues, we propose a novel framework GR-Gaussian, which introduces graph structures to model relationships between neighboring points. Specifically, we first introduce a \textbf{Denoised Point Cloud Initialization Strategy(De-Init)} based on an enhanced FDK method, which reduces noise and artifacts while preserving structural details. Then, we develop a \textbf{Pixel-Graph-Aware Gradient Strategy(PGA)} that reconstructs the gradient computation process by incorporating inter-point information. This strategy enables more accurate splitting decisions for Gaussian kernels, effectively mitigating needle-like artifacts and enhancing spatial coherence. Finally, we design an efficient optimization strategy tailored for GR-Gaussian, ensuring robust and accurate reconstruction.

We validate our framework through extensive experiments on both simulated X-3D dataset\cite{r2_gaussian} and Real-world CT datasets \cite{FIPS_CT_dataset}, comparing it against multiple baseline methods. The results demonstrate that GR-Gaussian outperforms existing approaches in reconstruction quality, achieving PSNR gains of \textbf{0.67 dB} and \textbf{0.92 dB}, and SSIM improvements of \textbf{0.011} and \textbf{0.021}. Our main contributions are summarized as follows:
\begin{itemize}
    \item We propose a novel framework GR-Gaussian, which leverages graph-structured relationships to model objects, providing a solution for sparse-view tomographic reconstruction.
    \item To address needle-like artifacts in 3DGS-based methods under sparse-view conditions, we introduce the De-Init to enhance the quality of initialization. Then we propose the PGA to improve the gradient computation process.
    \item Our framework achieves state-of-the-art performance on the X-3D and real-world datasets, demonstrating its effectiveness under challenging sparse-view conditions.
\end{itemize}

\begin{figure*}[tb]
  \centering
  \includegraphics[width=\linewidth]{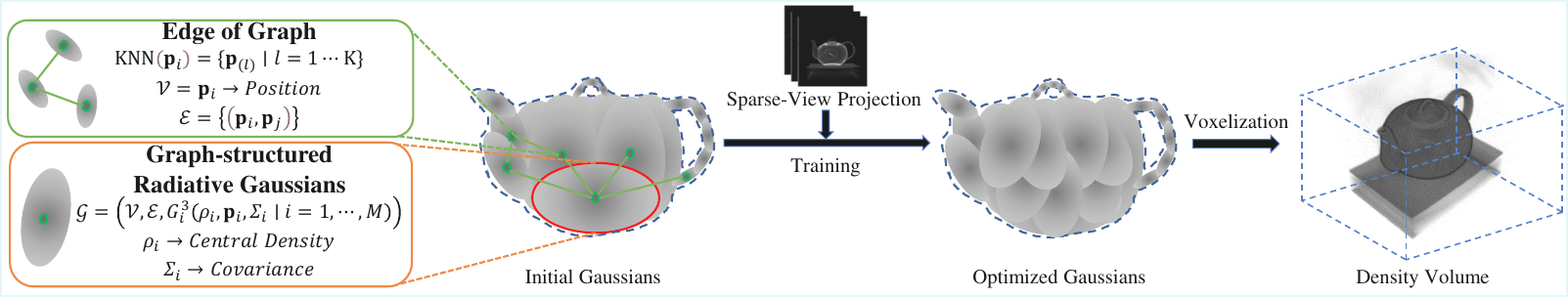}
  \caption{The scanned object is represented as \textbf{graph-based radiative Gaussians}, optimized using real X-ray projections to retrieve the density volume via voxelization.} 
  \label{Framework}
\end{figure*}

\section{Related work}
\subsection{Tomographic Reconstruction}
Computed Tomography (CT) is a critical imaging technique widely used in medicine\cite{hounsfield1980computed,katsuragawa2007computer} and industry\cite{de2014industrial,zang2019warp}. Sparse-view CT reconstruction, where the number of projections is limited, poses significant challenges for traditional methods. Traditional analytical methods, such as Filtered Back Projection (FBP) and its 3D counterpart FDK\cite{feldkamp1984practical}, are computationally efficient but prone to severe streak artifacts in sparse-view scenarios, leading to degraded image quality. Iterative methods\cite{andersen1984simultaneous,sidky2008image,manglos1995transmission,sauer2002local}, which optimize reconstruction through regularized energy minimization, can suppress artifacts but are computationally expensive and risk losing fine structural details. Deep learning methods\cite{lin2023learning,chung2023solving,liu2023dolce} have shown promise in CT reconstruction, particularly in sparse-view scenarios. Supervised approaches leverage large datasets to learn semantic priors for tasks like projection inpainting\cite{anirudh2018lose,ghani2018deep} and volume denoising\cite{chung2023solving,lee2023improving,liu2023dolce,liu2020tomogan}, but struggle with generalization to unseen data. Self-supervised methods, inspired by NeRF\cite{mildenhall2021nerf}, optimize density fields using photometric losses\cite{cai2024structure,zha2022naf,zang2021intratomo} but are computationally prohibitive due to extensive point sampling during rendering.
\subsection{3D Gaussian Splatting}
3D Gaussian Splatting has demonstrated remarkable performance in RGB tasks, such as surface reconstruction\cite{Li2025mmwbg,Lin2025davgs,Hou2025ls3dgs}, dynamic scene modeling\cite{zhu2025liag,Huang2025FatesGS,wu20244d}, human avatar creation\cite{Qu2025HOGSA,li2024animatable}, and 3D generation, outperforming NeRF in terms of speed by leveraging highly parallelized rasterization for image rendering. Recent studies have attempted to extend 3DGS to X-ray imaging. For example, Cai\cite{cai2024radiative} empirically modified 3DGS to synthesize novel-view X-ray projections, primarily using it as a data augmentation tool for traditional algorithms rather than directly generating CT models. Zha\cite{r2_gaussian} represented the density field with customized Gaussian kernels, yet they replaced the efficient rasterization with existing CT simulators. While effective in certain cases, their approach overfits sparse-view projections, leading to reconstruction failures.
\section{Method}
In this section, we present GR-Gaussian, a novel framework that represents objects through graph-structured relationships (Sec. 3.2). Specifically optimized for sparse-view CT reconstruction, our framework effectively tackles critical issues, most notably the emergence of needle-like artifacts (Sec. 3.3). We first introduce a Denoised Point Cloud Initialization Strategy based on an enhanced FDK method. Subsequently, a Pixel-Graph-Aware Gradient Strategy is proposed to refine the gradient computation process. Finally, we detail an optimization strategy specifically tailored for GR-Gaussian.

\subsection{Preliminaries}
\subsubsection{Radiative Gaussians}

To represent the target object, previous approach\cite{r2_gaussian} utilize a set of learnable 3D Gaussian kernels, denoted as $\mathbb{G}^3 = \{ G_i^3 \}_{i=1,\cdots,M}$, which we refer to as Radiative Gaussians. Each kernel $G_i^3$ defines a local Gaussian-shaped density field, expressed as:
\begin{equation}
 G_i^3(\mathbf{x} \mid \rho_i, \mathbf{p}_i, \Sigma_i) = \rho_i \cdot e^{\left(-\frac{1}{2} (\mathbf{x} - \mathbf{p}_i)^\top \Sigma_i^{-1} (\mathbf{x} - \mathbf{p}_i)\right) }
\end{equation}
Here, $\rho_i$ represents the central density, $\mathbf{p}_i \in \mathbb{R}^3$ denotes the position, and $\Sigma_i \in \mathbb{R}^{3 \times 3}$ is the covariance matrix, all of which are learnable parameters. For optimization, \( \Sigma_i \) is decomposed as \( \Sigma_i = \mathbf{R}_i \mathbf{S}_i \mathbf{S}_i^\top \mathbf{R}_i^\top \), where \( \mathbf{R}_i \) and \( \mathbf{S}_i \) are the rotation and scale matrices, respectively. The overall density at a given position $\mathbf{x} \in \mathbb{R}^3$ is computed as the sum of the contributions from all kernels:
\begin{equation}
\sigma(\mathbf{x}) = \sum_{i=1}^{M} G_i^3(\mathbf{x} \mid \rho_i, \mathbf{p}_i, \Sigma_i) 
\end{equation}
Notably, this formulation omits view-dependent color since X-ray attenuation depends solely on isotropic density, rendering it ideal for CT reconstruction.

\subsubsection{Pixel-Aware Gradient}
In 3D Gaussian Splatting, the decision to split or clone a Gaussian kernel \( G_i^3 \) is based on the average gradient magnitude of its Normalized Device Coordinates (NDC) across multiple viewpoints. For viewpoint \( v \), the NDC coordinates are \(\mu_{\mathrm{ndc}}^{i,v} = (\mu_{\mathrm{ndc,x}}^{i,v}, \mu_{\mathrm{ndc,y}}^{i,v})\), and the corresponding loss is \( L_v \). An Adaptive Density Control (ADC) mechanism adjusts kernel density every 100 iterations. A splitting or cloning operation is triggered if the average gradient magnitude across \( N^i \) viewpoints exceeds a threshold \( \tau_{\mathrm{pos}} \):
\begin{equation}
\frac{1}{N^i}\sum_{v=1}^{N^i}\left\|\frac{\partial L_v}{\partial\mu_{\mathrm{ndc}}^{i,v}}\right\|_2 > \tau_{\mathrm{pos}},
\end{equation}

During photometric reconstruction tasks, the gradient flow through color channels drives density field optimization. For a Gaussian kernel \( G_i^3 \) at viewpoint \( v \), the gradient contribution is computed as:
\begin{equation}
(\mathbf{g}_i^{r})^v = \frac{\partial L_v}{\partial\mu_{\mathrm{ndc}}^{i,v}} = \sum_{pix=1}^{m_v^i}\sum_{j=1}^{3} \frac{\partial L_v}{\partial c_j^{pix}} \cdot \frac{\partial c_j^{pix}}{\partial \alpha_{v,pix}^i} \cdot \frac{\partial \alpha_{v,pix}^i}{\partial\mu_{\mathrm{ndc}}^{i,v}},
\end{equation}
Here, \( \alpha_{v,pix}^i \) represents the contribution factor of a pixel to \( G_i^3 \), \( c_j^{pix} \) is the intensity of the \( j \)-th color channel, and \( m_v^i \) is the total number of pixels involved. \( \alpha_{v,pix}^i \) is defined as a function of the Euclidean distance between the Gaussian's center and the pixel center, decaying exponentially with increasing distance.

Within CT reconstruction tasks, where images are grayscale, gradient computation is simplified by removing the summation over color channels. The gradient contribution of a pixel under viewpoint \(v\) to Gaussian \(G_i^3\) is:
\begin{equation}
(\mathbf{g}_i^{c'})^v = \sum_{pix=1}^{m_v^i} \frac{\partial L_v}{\partial \alpha_{v,pix}^i} \cdot \frac{\partial \alpha_{v,pix}^i}{\partial \mu_{\mathrm{ndc}}^{i,v}}.
\end{equation}
This adaptation focuses on density contributions and ignores color information. The exponential decay of \( \alpha_{v,pix}^i \) emphasizes the influence of pixels closer to the Gaussian center, enabling precise gradient updates.

\subsection{Representing Objects with Graph-based Radiative Gaussians}
As shown in Figure \ref{Framework}, an object is represented as a learnable 3D Gaussian graph \(\mathcal{G}\), defined as:
\begin{equation} \label{eq:graph_definition}
\mathcal{G} = (\mathcal{V}, \mathcal{E}, \{G_i^3(\rho_i, \mathbf{p}_i, \Sigma_i)\}_{i=1}^M),
\end{equation}
where \(\mathcal{V} = \{\mathbf{p}_i\}_{i=1}^M\) denotes the positions of the Gaussian kernels, \(\mathcal{E}\) represents the edges connecting the kernels, and \(G_i^3\) is a Gaussian kernel parameterized by density \(\rho_i\), position \(\mathbf{p}_i \in \mathbb{R}^3\), and covariance \(\Sigma_i \in \mathbb{R}^{3 \times 3}\).

The graph structure \(\mathcal{G}\) is constructed to capture spatial relationships between Gaussian kernels. The vertices \(\mathcal{V}\) represent kernel positions, while the edges \(\mathcal{E}\) are determined using the KNN algorithm:
\begin{equation} \label{eq:knn_definition}
\text{KNN}(\mathbf{p}_i) = \{\mathbf{p}_j \mid j \in \text{argmin}_{j} \, d(\mathbf{p}_i, \mathbf{p}_j), \, j = 1, \cdots, K\},
\end{equation}
where the Euclidean distance \(d(\mathbf{p}_i, \mathbf{p}_j)\) is given by:
\begin{equation} \label{eq:euclidean_distance}
d(\mathbf{p}_i, \mathbf{p}_j) = \|\mathbf{p}_i - \mathbf{p}_j\|_2.
\end{equation}
To ensure bidirectional connectivity, the edge set \(\mathcal{E}\) is defined as:
\begin{equation} \label{eq:edge_set}
\mathcal{E} = \{(\mathbf{p}_i, \mathbf{p}_j) \mid \mathbf{p}_j \in \text{KNN}(\mathbf{p}_i) \text{ and } \mathbf{p}_i \in \text{KNN}(\mathbf{p}_j)\}.
\end{equation}

Each Gaussian kernel \(G_i^3\) defines a local density field by combining its own density with contributions from neighboring kernels:
\begin{equation} \label{eq:local_density}
G_i^3(\mathbf{x}) = \rho_i \cdot G(\mathbf{x} \mid \mathbf{p}_i, \Sigma_i) + \sum_{j \in \mathcal{N}(i)} w_{ij} \cdot \rho_j \cdot G(\mathbf{x} \mid \mathbf{p}_j, \Sigma_j),
\end{equation}
where \(G(\mathbf{x} \mid \mathbf{p}, \Sigma)\) is the Gaussian function:
\begin{equation} \label{eq:gaussian_function}
G(\mathbf{x} \mid \mathbf{p}, \Sigma) = \exp\left(-\frac{1}{2} (\mathbf{x} - \mathbf{p})^\top \Sigma^{-1} (\mathbf{x} - \mathbf{p})\right),
\end{equation}
The weight \(w_{ij}\) is introduced to quantify the influence of neighboring kernels based on their spatial proximity. It is defined as:
\begin{equation} \label{eq:weight_definition}
w_{ij} = \exp\left(-\frac{\|\mathbf{p}_i - \mathbf{p}_j\|^2}{k}\right),
\end{equation}
where \(\|\mathbf{p}_i - \mathbf{p}_j\|^2\) represents the squared Euclidean distance between kernel positions \(\mathbf{p}_i\) and \(\mathbf{p}_j\), and \(k\) is a scaling factor that controls the sensitivity of the weight to distance. The combined density \(D(\mathbf{x})\) at a position \(\mathbf{x}\) is computed as:
\begin{equation}
\label{eq:combined_density}
D(\mathbf{x}) = \sum_{i=1}^M G_i^3(\mathbf{x}).
\end{equation}

This graph-based representation is specifically designed to model the relationships between neighboring points, enabling the incorporation of inter-point information. By leveraging these relationships, the framework facilitates the optimization of update strategies, improving the accuracy of the reconstruction process.

\subsection{Training Graph-based Radiative Gaussians}

Our training pipeline, illustrated in Figure \ref{pipeline}, begins with the initialization of Graph-based Radiative Gaussians from a modified FDK volume. A graph is then constructed using the KNN method to capture spatial relationships between Gaussian kernels. Projections are rasterized to compute photometric losses, while tiny density volumes are voxelized for 3D regularization. Finally, a modified Pixel-Graph-Aware Gradient Strategy is applied to densify the Gaussians.

\subsubsection{Denoised Point Cloud Initialization Strategy}

Previous methods utilize FDK to generate low-quality volumes for initialization. However, under sparse-view conditions, FDK reconstructions suffer from significant noise and artifacts, which degrade subsequent processing. To address this, as shown in Figure \ref{pipeline} (b), we propose a Denoised Point Cloud Initialization Strategy leveraging 3DGS characteristics. Gaussian filtering is applied to the FDK-reconstructed point cloud to suppress noise and artifacts while preserving structural details, ensuring high-quality initialization for robust and accurate reconstruction. The Gaussian filtering process is defined as:
\begin{equation}
\label{eq:gaussian_filter}
 f'(i) = \frac{1}{Z} \sum_{j \in \mathcal{N}(i)} G(j; \sigma_d) \cdot f(i + j), .
\end{equation}
Here, \(i\) denotes the coordinates of the current voxel, and \(j\) represents the offset within its neighborhood \(\mathcal{N}(i)\), defined as \(j_k \in [-r, r]\) for all dimensions \(k\). The Gaussian kernel is given by:
\[
G(j; \sigma_d) = \exp\left(-\frac{|j|^2}{2\sigma_d^2}\right),
\]
where \(|j|^2 = j_1^2 + j_2^2 + \cdots + j_n^2\), and \(Z = \sum_{j \in \mathcal{N}(i)} G(j; \sigma_d)\) is the normalization factor ensuring the kernel weights sum to 1. Empty regions are excluded using a density threshold \(\tau\), and \(M\) points are randomly sampled as kernel positions. Following \cite{r2_gaussian}, Gaussian scales are set based on nearest neighbor distances, assuming no rotation. Central densities are directly queried from the denoised volume.

\subsubsection{Pixel-Graph-Aware Gradient Strategy}
Previous methods compute gradients based solely on average values, limiting the contribution of large kernels with small gradients to the densification process. To address this limitation, we propose a Pixel-Graph-Aware Gradient Strategy that integrates graph-based relationships to refine gradient computation and enhance the splitting process. This strategy leverages the density differences between Gaussian kernels and their neighbors, consistent with the characteristics of X-ray CT imaging, where similar tissues or materials exhibit approximately constant attenuation coefficients. The density difference between a Gaussian kernel \(i\) and its neighbors \(\mathcal{N}(i)\) is defined as:
\begin{equation}
\Delta \rho_{ij} = |\rho_i - \rho_j|, \quad \forall j \in \mathcal{N}(i).
\end{equation}
The density difference \( \Delta \rho_{ij} \) between a Gaussian kernel \( G_i^3 \) and its neighbor \( G_j^3 \) reflects significant variations, with larger values contributing to higher gradient magnitudes. To enhance gradient computation, we incorporate these density differences into the gradient formula. For a Gaussian kernel \( G_i^3 \), the augmented gradient is defined as:
\begin{equation}
(\mathbf{g}_i^c)^v =
\sum_{pix=1}^{m_v^i} \frac{\partial L_v}{\partial \alpha_{v,pix}^i} \cdot
\frac{\partial \alpha_{v,pix}^i}{\partial\mu_{\mathrm{ndc}}^{i,v}} + \lambda_{g} \cdot \frac{\sum_{j \in \mathcal{N}(i)}  \Delta \rho_{ij}}{k},
\end{equation}
where \( (\mathbf{g}_i^c)^v \) denotes the augmented gradient under viewpoint \( v \), \( \lambda_g \) is a regularization parameter controlling the influence of neighboring kernels, and \( \mathcal{N}(i) \) represents the set of neighbors connected to \( G_i^3 \) in the graph. By leveraging graph-based relationships, the augmented gradient effectively captures density variations:
\begin{equation}
(\mathbf{g}_i^c)^v \propto \Delta \rho_{ij}, \quad \forall j \in \mathcal{N}(i).
\end{equation}
This approach enhances gradient estimation for large kernels, enabling more effective splitting. The increased gradient magnitude improves the representation of the underlying density field, governed by the splitting condition \( \|(\mathbf{g}_i^c)^v\| > \tau_{\mathrm{pos}} \), where \( \tau_{\mathrm{pos}} \) is a predefined threshold. The Pixel-Graph-Aware Gradient Strategy integrates spatial relationships into gradient computation for 3D Gaussian Splatting, leveraging graph-based density differences to refine the splitting strategy. 


\begin{figure}[h]
\centering
\includegraphics[width=\linewidth]{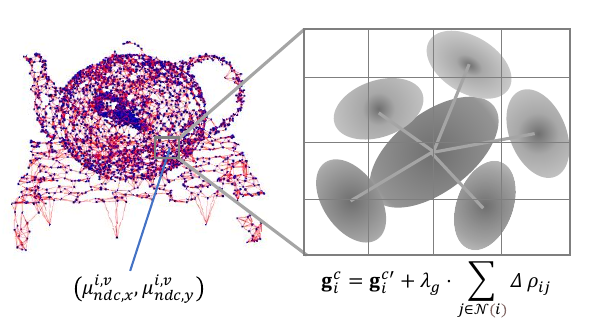}
\caption{The \textbf{Pixel-Graph-Aware Gradient Strategy} leverages density differences between Gaussian kernels by constructing a graph to encode point-to-point relationships, enhancing gradient computation and enabling effective splitting of large kernels with low gradients. 
    }

    \label{reason}
\end{figure}

\subsubsection{Density Voxelizer}

We employ a voxelizer, denoted as $ \mathbf{V} $, to efficiently extract a density volume $V \in R^{X \times Y \times Z} $ from the graph-based radiative representation, defined as $V = \mathbf{V}(\mathcal{G})$. The voxelizer \cite{r2_gaussian} divides the target space into $ 8 \times 8 \times 8 $ 3D tiles and performs Gaussian culling to retain kernels with a 99\% confidence of intersecting each tile. Building on this, we incorporate graph-based relationships into the voxelization process. The final voxel values are computed in parallel, as detailed in Eq.~\ref{eq:combined_density}.

\subsubsection{Optimization}
To optimize the radiative Gaussians, we employ stochastic gradient descent. The total loss function consists of the photometric loss \( \mathcal{L}_1 \), the D-SSIM loss \( \mathcal{L}_{ssim} \)\cite{wang2004image}, the 3D total variation (TV) regularization \( \mathcal{L}_{tv} \), and the graph Laplacian regularization \( \mathcal{L}_{lap} \). The graph Laplacian regularization term \( \mathcal{L}_{lap} \) is defined as:
\begin{equation}
\mathcal{L}_{lap}(\mathcal{G}) =  \sum_{i=1}^M \sum_{j \in \mathcal{N}(i)} w_{ij} (\rho_i - \rho_j)^2,
\end{equation}
where \( \mathcal{G} \) represents the graph structure of the Gaussian kernels, \( \rho_i \) and \( \rho_j \) denote the densities of Gaussian kernels \( G_i^3 \) and \( G_j^3 \), and \( w_{ij} \) is the weight of the edge \( (i, j) \). This term encourages local smoothness by minimizing density differences between neighboring kernels while preserving boundary information. To further enhance regularization, we define \( \mathcal{L}_{norm} \) as:
\begin{equation}
\mathcal{L}_{norm} =  \lambda_{lap}\mathcal{L}_{lap}(\mathcal{G}) + \lambda_{tv}\mathcal{L}_{tv}(\mathcal{V}_{tv}) ,
\end{equation}
where \( \mathcal{L}_{tv} \) represents the 3D total variation regularization applied to the volume \( \mathcal{V}_{tv} \). The total loss function is then expressed as:
\begin{equation}
\mathcal{L}_{total} = \mathcal{L}_1(\mathbf{I}_r, \mathbf{I}_m) + \lambda_{ssim} \mathcal{L}_{ssim}(\mathbf{I}_r, \mathbf{I}_m) + \mathcal{L}_{norm}.
\end{equation}
Here, \( \mathbf{I}_r \) and \( \mathbf{I}_m \) denote the reconstructed and measured images, respectively.
\begin{table*}[ht]
\centering
\tabcolsep=6pt
\begin{tabular}{ccccccccc|cc}
\hline
\multirow{2}{*}{\textbf{Method}}& \multicolumn{2}{c}{\textbf{HO}} & \multicolumn{2}{c}{\textbf{BS}} & \multicolumn{2}{c}{\textbf{AO}} &\multicolumn{2}{c}{\textbf{Average}} &\multicolumn{2}{c}{\textbf{RD}}   \\ 
\cmidrule(r){2-3} \cmidrule(r){4-5} \cmidrule(r){6-7} \cmidrule(r){8-9} \cmidrule(r){10-11}
& \textbf{PSNR} & \textbf{SSIM} &\textbf{PSNR} &\textbf{SSIM}& \textbf{PSNR} & \textbf{SSIM }  &\textbf{PSNR} & \textbf{SSIM } &\textbf{PSNR} & \textbf{SSIM } \\ 
\hline
FDK & 22.64 & 0.319  & 23.45 & 0.296  & 22.95 & 0.336 & 23.01 & 0.317 & 23.30 & 0.335  \\ 
SART& 29.58 & 0.773  & 32.37 & 0.878  & 31.48 & 0.825 & 31.14 & 0.825 & 31.52 & 0.790  \\ 
ASD-POCS& 29.42 & 0.810  & 31.40 & 0.887  & 30.58 & 0.845 & 30.47 & 0.847 & 31.32 & 0.810 \\ 
NAF       & 32.05 & 0.841  & 34.36 & 0.930  & 35.34 & 0.909 & 33.92 & 0.893 & 32.92 & 0.772  \\ 
SAX-NeRF  & 32.53 & 0.858 & 34.67 & 0.940 & 35.85 & 0.917  & 34.35 & 0.905 & 33.49 & 0.793 \\ 
$R^2$-GS & 32.98 & 0.881 & 35.08 & 0.944 & 37.52 & 0.945 & 35.19 & 0.922 & 35.03 & 0.837 \\ 
Ours & \textbf{33.47} & \textbf{0.891} & \textbf{35.56} & \textbf{0.952} & \textbf{38.55} & \textbf{0.955} & \textbf{35.86} & \textbf{0.933}  & \textbf{35.95} & \textbf{0.858}\\ 
\hline
\end{tabular}
\caption{\textbf{Detailed quantitative results} under 25-view. Comparison of our GR-Gaussian with different methods on the X-3D dataset and Real-world dataset. HO: Human Organs, BS: Biological Specimens, AO: Artificial Objects (all from the X-3D dataset); RD: Real-world Dataset.}
\label{table1}
\end{table*}

\section{Experiments}
\subsection{Experimental Settings}
\subsubsection{Dataset}
Following $R^2$-Gaussian , we utilize the publicly available X-3D dataset to evaluate our method. This dataset\cite{r2_gaussian} encompasses diverse categories, including human organs(chest, foot, head, jaw, and pancreas), artificial objects(backpack, engine, mount, present, and teapot), and biological specimens(beetle, bonsai, broccoli, kingsnake, and pepper), providing a comprehensive benchmark for reconstruction performance. Using the tomography toolbox TIGRE, we generate $512 \times 512$ projections along a full circular trajectory, incorporating electric noise and ponton scatter to simulate realistic conditions. Subsequently, CT volumes are scanned within an angular range of $0^\circ$ to $360^\circ$ to produce 25-view projections. For the Real-World datasets\cite{FIPS_CT_dataset}, the projections are directly acquired using CT scanners and correspond to specific angles. Each projection has a resolution of $560 \times 560$. To achieve sparse-view conditions, we uniformly select 25 projection angles from the full angular range of $0^\circ$ to $360^\circ$. This dataset including pine, seashell and walnut.

\subsubsection{Implementation Details}
The GR-Gaussian framework is implemented using PyTorch\cite{paszke2019pytorch}, PyTorch Geometric\cite{fey2019fast}, and CUDA\cite{guide2013cuda}, and optimized with the Adam optimizer\cite{diederik2014adam}. Initial learning rates for location, density, scale, and rotation are set to 0.0002, 0.01, 0.005, and 0.001, respectively, with exponential decay reducing them to 10\% of their initial values. Regularization parameters include a television volume level of $D = 32$, loss weights $\lambda_{\mathrm{ssim}} = 0.25$ and $\lambda_{\mathrm{tv}} = 0.05$, a Graph Laplacian weight $\mathcal{L}_{\mathrm{lap}} = 8 \times 10^{-4}$, a Gaussian denoising parameter $\sigma_d = 3$, and a gradient computation weight $\lambda_{\mathrm{g}} = 1 \times 10^{-4}$. The framework uses $M = 50,000$ Gaussians, a density threshold $\tau = 0.001$, and $k = 6$ nearest neighbors. A dynamic stopping criterion ($Iter_{stop}$) evaluates PSNR every 500 iterations, terminating if PSNR decreases by more than 0.5\%. All experiments were conducted on an RTX 4090 GPU. Quantitative evaluation employed PSNR and SSIM\cite{wang2004image}, with PSNR assessing 3D reconstruction accuracy and SSIM evaluating structural consistency across 2D slices.

\begin{figure*}[t]
\centering
\includegraphics[width=0.8\textwidth]{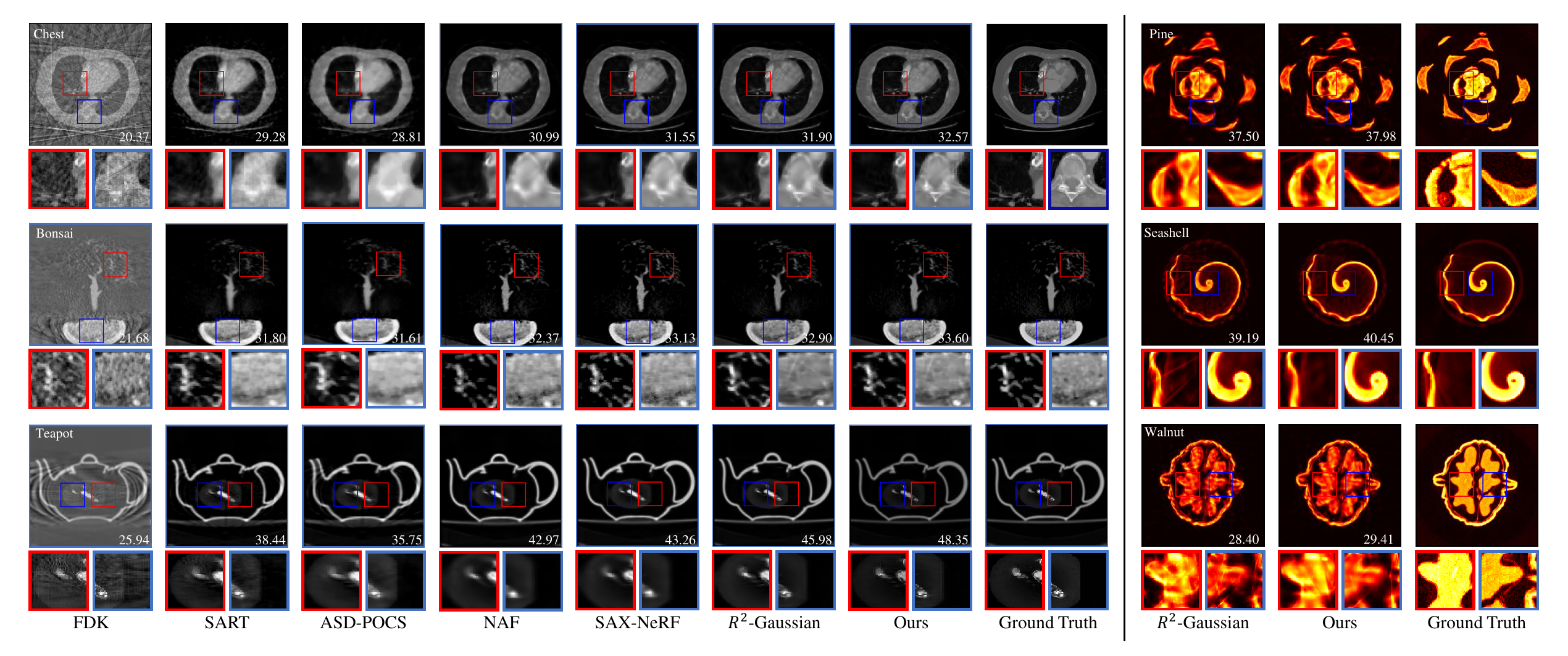} 
\caption{The left shows \textbf{CT reconstructions from the X-3D dataset}, covering three categories: chest, bonsai, and teapot. The right displays \textbf{Real-world dataset reconstructions} with colorized slices to highlight details, all under 25-view conditions.}
\label{result}
\end{figure*}

\subsection{Results and Evaluation}
The experiments aim to evaluate the performance of GR-Gaussian in sparse-view tomographic reconstruction, focusing on reconstruction quality across X-3D and Real-world datasets. On the X-3D dataset, GR-Gaussian outperforms the SOTA 3DGS-based method \cite{r2_gaussian}, NeRF-based methods \cite{zha2022naf,cai2024structure} and three traditional methods \cite{feldkamp1984practical,andersen1984simultaneous,sidky2008image} in both PSNR and SSIM. As show in Table \ref{table1}, it achieves a PSNR increase of \textbf{0.67 dB} and an SSIM improvement of \textbf{0.011}, demonstrating its ability to suppress artifacts and enhance structural consistency. In the experiments with real-world datasets, GR-Gaussian demonstrates significant robustness against noise and artifacts, achieving a PSNR improvement of \textbf{0.92 dB} and an SSIM improvement of \textbf{0.021}. These results highlight the method's adaptability to practical scenarios, where data imperfections pose significant challenges to reconstruction quality. As shown in Figure \ref{result}, GR-Gaussian produces visually superior reconstructions with sharper edges and fewer streak artifacts compared to other methods. These qualitative improvements align with the quantitative results, further validating the effectiveness of the proposed framework.

\begin{table}[t]
\centering
\tabcolsep=2pt
\fontsize{8pt}{8pt}\selectfont
\begin{tabular}{ccc|cccc}
  \hline
\multirow{2}{*}{\textbf{Baseline}} & \multirow{2}{*}{\textbf{De-Init}} & \multirow{2}{*}{\textbf{PGA}} & \multicolumn{2}{c}{\textbf{X-3D}}& \multicolumn{2}{c}{\textbf{Real-world}}\\
&    &    & \textbf{PSNR} & \textbf{SSIM}& \textbf{PSNR} & \textbf{SSIM} \\ \hline
\checkmark &        &      & 35.20 & 0.923  & 35.28 & 0.836 \\ 
\checkmark& \checkmark     &      & 35.61 & 0.931  & 35.90 & 0.857 \\ 
\checkmark &        & \checkmark    & 35.77 & 0.932  & 35.62 & 0.848 \\ 
\checkmark & \checkmark      & \checkmark    & 35.86 & 0.932  & 35.95 & 0.858 \\ \hline
\end{tabular}
\caption{\textbf{Ablation study} of the components of the GR-Gaussian on the X-3D dataset and Real-world dataset.}
\label{table2}
\end{table}

\begin{figure}[t]
\centering
\includegraphics[width=0.9\columnwidth]{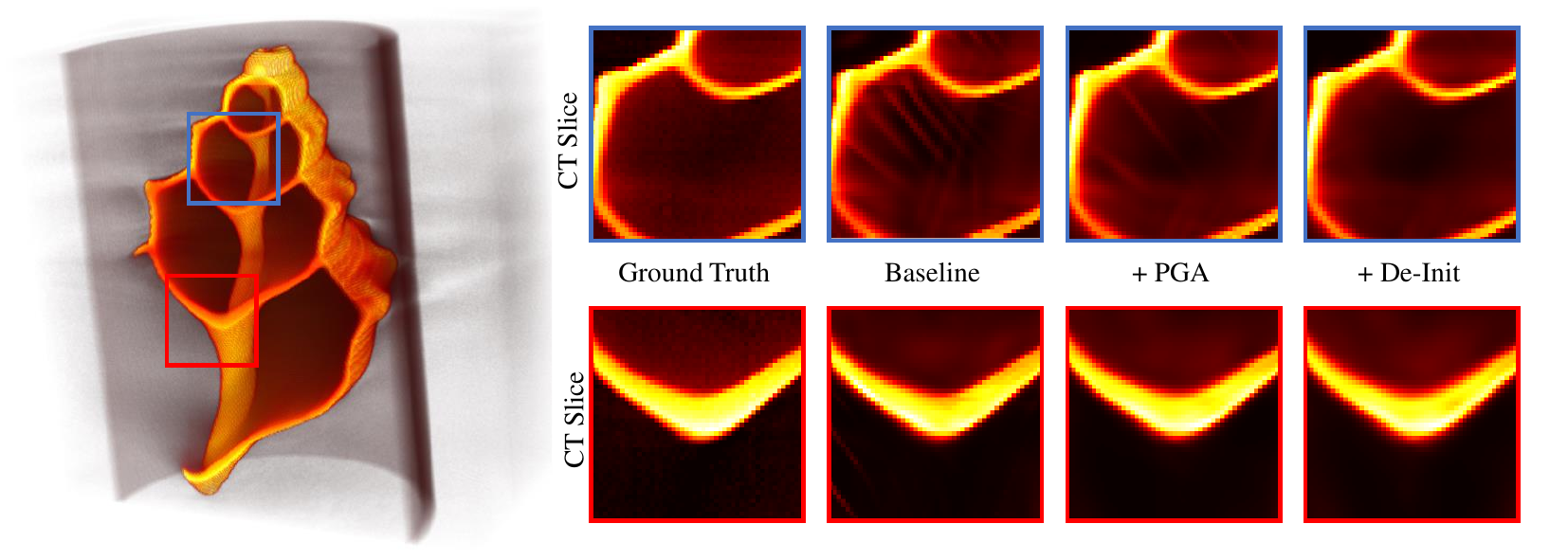} 
\caption{\textbf{Ablation study} results highlight the impact of PGA and De-Init in enhancing reconstruction quality.}
\label{Component analysis visual}
\end{figure}

\subsection{Ablation Study}
\subsubsection{Component Analysis}
We performed ablation experiments to assess the contributions of Denoised Point Cloud Initialization (De-Init) and Pixel-Graph-Aware Gradient Strategy (PGA) to reconstruction performance on both the simulated X-3D dataset and the real-world dataset. For the ablation study, the baseline model omits the De-Init and PGA components, with Gaussians initialized via an FDK-based approach. To validate the component's efficiency, we measure PSNR and SSIM under 25 views, with results listed in Table \ref{table2}. Additionally, Figure \ref{Component analysis visual} presents colorized volume and slice examples to visually highlight detailed improvements. Across the Real-world datasets, De-Init effectively mitigates needle-like artifacts, while PGA exhibits superior performance in preserving smooth regions.

\begin{table}[h]
  \centering
  \tabcolsep=2pt
  \fontsize{8pt}{8pt}\selectfont
  \caption{Quantitative results for different k-nearest neighbors and $\sigma_d$. Analysis of the X-3D dataset and Real-world dataset.}
  {
      \begin{tabular}{l|c c c c c}
  \toprule
      \textbf{Metric} & \textbf{k=4} & \textbf{k=5} & \textbf{k=6} & \textbf{k=7} & \textbf{k=8} \\
      \midrule \midrule
      \multicolumn{6}{c}{\textbf{X-3D Dataset}} \\
      \midrule
      PSNR & 35.80 & 35.84 & 35.86 & 35.87 & 35.89 \\
      SSIM & 0.932 & 0.932 & 0.933 & 0.933 & 0.934 \\
      Time & 536s & 567s & 629s & 763s & 865s \\
      \midrule
      \multicolumn{6}{c}{\textbf{Real World Dataset}} \\
      \midrule
      PSNR & 35.82 & 35.87 & 35.95 & 35.97 & 36.01 \\
      SSIM & 0.851 & 0.854 & 0.858 & 0.858 & 0.860 \\
      Time & 562s & 597s & 669s & 794s & 927s \\
      \midrule
      \textbf{$\sigma_d$} & \textbf{1} & \textbf{2} & \textbf{3} & \textbf{4} & \textbf{5} \\
      \midrule
      \multicolumn{6}{c}{\textbf{X-3D Dataset}} \\
      \midrule
      PSNR & 35.60 & 35.72 & 35.86 & 35.68 & 35.43 \\
      SSIM & 0.926 & 0.928 & 0.933 & 0.927 & 0.925 \\
      \midrule
      \multicolumn{6}{c}{\textbf{Real World Dataset}} \\
      \midrule
      PSNR & 35.90 & 35.93 & 35.95 & 35.91 & 35.85 \\
      SSIM & 0.857 & 0.858 & 0.858 & 0.857 & 0.856 \\
  \bottomrule
      \end{tabular}
  }
  \label{parameter_analysis}
  \end{table}

\subsubsection{Parameter Analysis}
The number of neighbors $k$ and the standard deviation $\sigma_d$ of the Gaussian kernel are critical hyperparameters in our method, influencing spatial relationship modeling and noise suppression. Sensitivity analysis on the X-3D and real-world datasets (Table \ref{parameter_analysis}) shows optimal reconstruction quality at $k=6$ and $\sigma_d=3$, achieving the highest PSNR and SSIM. Larger $k$ improves gradient accuracy by capturing richer neighborhoods but increases computational complexity, while smaller $k$ reduces overhead but compromises gradient precision. For De-Init, $\sigma_d$ directly affects filtering efficacy: smaller $\sigma_d$ preserves local details but inadequately suppresses noise, whereas larger $\sigma_d$ smooths noise effectively but risks over-smoothing edges and details.

\begin{table}[h]
\centering
  \tabcolsep=2pt
  \caption{Quantitative results of SSGU extension.} 
  \fontsize{8pt}{8pt}\selectfont
{
    \begin{tabular}{l|c c c }
\toprule
    \textbf{Method} & \textbf{w/o SSGU} & \textbf{w/ SSGU $\lambda=3$} & \textbf{w/ SSGU $\lambda=5$} \\
    \midrule \midrule
    SDS & 40.37s & 14.12s & 8.62s \\
    DDS & 66.89s & 22.65s & 13.90s \\
    CoSD & 344.76s & 128.97s & 79.15s \\
\bottomrule
    \end{tabular}
    }
    \label{tab:SSGU extension}
\end{table}

\begin{figure}[t]
\centering
\includegraphics[width=0.9\columnwidth]{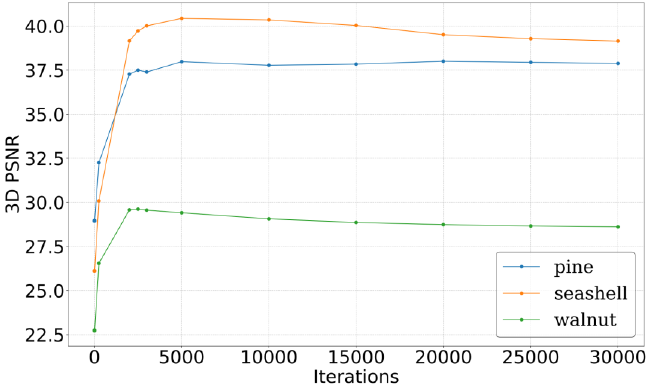} 
\caption{\textbf{Iteration Analysis} Reconstruction results of GR-Gaussian across different iterations, illustrating the impact of iteration count on PSNR.}
\label{iteration_analysis}
\end{figure}

\subsubsection{Iteration Analysis}
Under sparse-view conditions, PSNR initially improves with increased iterations but eventually declines (see Fig. \ref{iteration_analysis}). This reduction is less pronounced in the simulated X-3D dataset, yet it becomes evident in real-world datasets, where noise and artifacts exacerbate reconstruction degradation. To address this issue, we employ a dynamic iteration stopping criterion ($Iter_{stop}$), where PSNR is evaluated every 500 iterations. If the PSNR decreases by more than 0.5\%, the iteration process is terminated. This approach ensures that the model avoids overfitting and maintains optimal reconstruction quality.


\section{Conclusion}
This paper introduces GR-Gaussian, a novel framework based on 3D Gaussian Splatting for sparse-view tomographic reconstruction. Our approach employs a Denoised Point Cloud Initialization Strategy that robustly reduces noise while preserving critical structural details. Furthermore, by incorporating a graph-structured representation together with a novel gradient computation mechanism, GR-Gaussian significantly enhances reconstruction quality and effectively mitigates artifacts under challenging sparse-view conditions. Extensive evaluations on X-3D and real-world datasets demonstrate that our method outperforms existing techniques in both artifact suppression and reconstruction accuracy.
{
    \small
    \bibliographystyle{ieeenat_fullname}
    \bibliography{main}
}


\end{document}